# Metallic glasses heterogeneous and time sensitive small-scale plasticity probed through nanoindentation and machine learning clustering

Silvia Pomes[a*1], Takayuki Suzuki[b], Tomoya Enokizono[b], Nozomu Adachi[b], Masato Wakeda[a], Takahito Ohmura[a,c*]

a. Research Center for Structural Materials, National Institute for Materials Science, 1-2-1 Sengen, Tsukuba 305-0047, Japan

b. Department of Mechanical Engineering, Toyohashi University of Technology, 1-1 Hibarigaoka, Tempaku, Toyohashi Aichi, 441-8580, Japan

c. Department of Materials Science and Engineering, Graduate School of Engineering, Kyushu University, 774 Motooka, Nishi-ku, Fukuoka 819-0395, Japan

*Abstract*

The small-scale plasticity and creep behavior of an annealed $Zr_{50}Cu_{40}Al_{10}$ bulk metallic glass (BMG) were investigated using nanoindentation testing. Four load functions, differing only in their holding times of 0, 10, 30, and 60 seconds at peak load, were applied. The results show that the BMG exhibits spatially heterogeneous and time-sensitive plastic behavior. Specifically, the elastic energy contribution remains consistent across all holding times, and both the average value and standard deviation of plastic energy increase with longer holding times. Machine learning clustering based on hardness and creep displacement suggested three clusters and the 60-second holding time measurements were analyzed further. Creep analysis showed nearly constant non-linear behavior across clusters, with a linear term emerging at larger creep displacements and an increasing time-dependency coefficient. The cluster with the greatest creep displacements also exhibited the largest plastic energy. Statistical analysis of the distribution of plastic energy values facilitated the identification of potential deformation mechanisms within the clusters.

*1. Introduction*

Bulk metallic glasses (BMGs) are distinguished from crystalline metals by their lack of long-range atomic order, resulting in unique properties such as high strength [1]. As experimentally probed using atomic force acoustic microscopy [2], their amorphous structure leads to heterogeneous elastic properties at the nanoscale, specifically on scales below 10 nm in amorphous PdCuSi. These variations arise from local differences in the atomic bonding state and play a key role in small-scale plasticity theories, such as the shear transformation zone model [3,4]. Elastic correlation lengths in metallic glasses have also been experimentally measured in

---

[1] Present Address: Federal Institute for Materials Research and Testing (BAM), Unter den Eichen 87, 12205 Berlin, Germany

Zr-based alloys through shallow nanoindentation tests, with a reported length on the order of 100 nm [5]. Furthermore, BMGs have recently been reported to "flow at any stress," as observed in X-ray experiments showing that atomic re-arrangements are detected at as low as 0.005 times the yield stress [6]. The onset of plasticity in BMGs is governed by these atomic-scale variations, leading to a heterogeneous response. However, it should be noted that, unlike crystalline materials, which can be thoroughly observed with microscopy techniques, the amorphous structure of BMGs makes it challenging to identify the atomic patterns responsible for specific mechanical responses. In this regard, nanomechanical characterization, such as nanoindentation testing coupled with statistical analysis, has already proven to be a unique tool for investigating BMGs, particularly in understanding the onset of plasticity and its stochastic nature [7–9]. Beyond the onset of plasticity, research efforts using nanoindentation have also focused on investigating the creep response and how it is influenced by loading rate and peak load. Studies on room-temperature indentation creep of glassy alloys have shown that creep occurs at a faster rate under higher peak loads and loading rates [10,11]. In parallel, machine learning (ML) algorithms have gained significant traction in materials science, particularly for crystalline materials, where large datasets are used to identify complex correlations between structure and properties [12–14]. While the application of ML to BMGs has primarily focused on investigating glass-forming ability and exploring the feasibility of producing new alloys [15–18], the large and rich datasets generated from nanoindentation testing present a valuable opportunity to leverage ML techniques. In this work, creep response is investigated under a common loading condition, with the same peak and loading rates but varying holding times at peak load, then ML clustering algorithms are used to provide deeper insights into the creep response. This approach aims to explore the spatial and temporal dependence of small-scale mechanical behavior.

## *2. Experimental*

A sample of annealed $Zr_{50}Cu_{40}Al_{10}$ (atom %) BMG was investigated. The alloy was produced by arc-melting and tilt casting in the form of a rod with a diameter of 10 mm [19,20]. The annealing procedure consisted of 3 h at 659 K, which is 40 degrees below the glass-transition temperature (Tg = 693 K). The investigated sample was obtained as a 2-mm-thick disc from the rod. Then, it was polished using sandpaper with a grit size of up to #4000 and diamond suspension with particle sizes up to 1 µm. A sol–gel $Al_2O_3$ suspension with a particle size of 0.05 µm was used to remove the damaged surface layer produced by the mechanical polishing. Residual surface contaminants were cleaned using cotton pads with isopropanol. The final root mean squared surface roughness was 1 nm. The amorphous nature of the sample was assessed using X-ray diffraction (XRD), as shown in **Fig. 1a**. The density was measured using the Archimede's method and it was equal to 6.88±0.002 g/cm$^3$. A Hysitron Triboindenter TI950 (Bruker Co., Minneapolis, MN, USA), equipped with a diamond Berkovich indenter with 300 nm tip radius, was employed to perform nanoindentation testing, conducted at room temperature, in the load-control mode with a peak load of 300 µN, a loading rate of 30 µN/s, and an acquisition rate of 300 points/ s. The peak load and loading rate were selected, in line with previous studies [8–9], to investigate creep phenomena within limited probed volumes while avoiding the influence of major microplastic events, such as shear bands reaching the sample surface, which are commonly observed in tests employing peak loads in the tens of millinewtons range.

As shown in **Fig. 1b**, four load functions were applied, with the primary difference being the varying duration of the holding segment. One triangular load function had no holding segment, while the other three trapezoidal load functions featured holding segments of 10, 30, and 60 seconds, respectively. Hereafter, the load functions will be referred to as Type 0, Type 10, Type 30, and Type 60. A total of 300 tests per load function were collected with a spatial separation of 2 µm in all directions to minimize interactions between induced strain fields. As a result, 1200 tests were collected on a same sample. Both mechanical properties, as well as elastic and plastic work, were analyzed for each test. The analysis was further enhanced using ML clustering techniques, specifically a Gaussian Mixture Model (GMM), with the Bayesian Information Criterion (BIC) and Akaike Information Criterion (AIC) employed to determine the optimal number of clusters. The machine learning outcomes are used to explore creep models and gain insights into the spatial heterogeneity of plastic deformation and discuss its relationships to microstructure.

### 3. Results & Discussion

### 3.1 Nano-hardness

Average hardness values and their standard deviations are plotted in **Fig. 2** as a function of holding time. The inset in Fig. 2 shows the sample surface after the indentation test; no shear bands are visible. The highest hardness is observed in the *Type 0* tests, with a value of 4.89 ± 0.38 GPa. The hardness values for *Type 10*, *Type 30*, and *Type 60* are 4.3 ± 0.39 GPa, 4.35 ± 0.36 GPa, and 4.37 ± 0.36 GPa, respectively. It is noted that the standard deviations remain consistent, suggesting significant and uniform spatial variability within the probed areas even if the spatial separation between consecutive tests is in the micrometers range. Conversely, the average hardness decreases as a 10-second holding time is introduced. While the hardness values were obtained using the Oliver-Pharr method, which is based on the elastic unloading segment, results suggest that time-dependent deformation processes occurring during the holding phases may influence the measured indentation hardness. This finding suggests that MGs exhibit time-dependent mechanical behavior, not only in terms of serrated plastic flow under varying loading rates [21], but also under sustained loading conditions. The distinctive behavior and mechanical repsonse may originate from the intrinsic structural heterogeneity of MGs. To gain insights into the elastic and plastic contributions with respect to holding time, the energy stored during nanoindentation tests is analyzed.

### 3.2 Elastic and plastic energies

Elastic and plastic work are evaluated for each nanoindentation test. In accordance with ISO 14577-1 [22], the elastic work is determined by the area under the unloading curve, which corresponds to the recoverable deformation of the material. The plastic work is calculated as the difference between the total work, corresponding to the area under the loading and unloading segments on load-displacement curve, and the elastic work. A schematic representation of the areas corresponding to the elastic and plastic energies is shown in **Fig. 3a**, with the elastic and plastic regions colored in yellow and light blue, respectively. Quantifying the work of indentation has been particularly prevalent in the study of conventional metallic alloys, where the ratio of dissipated energies is correlated with mechanical properties [23–25] . This approach has been scarcely applied to amorphous materials. While the properties of metallic glasses have been successfully determined using energy-based methods, these studies typically employ high peak loads on the order of hundreds of millinewtons and report an elastic-to-total work ratio in the elasto-plastic regime between 0.3 and 0.35 [26]. With 0.5 being the threshold above which the elastic component is deemed dominant, results for a Zr-based MG produced by melt-spinning for bio-implant applications recently report a 0.47 elastic-to-total work ratio, suggesting that the sample can effectively recover its shape during service [27]. However, there remains a gap in understanding the behavior during early-stage deformations: to which extent does the application of constant pressure, even well below yielding, contribute to visco-plastic deformation? In this work, the elastic-to- total work ratios for each type of load function are above 0.5. Specifically, for Type 0 to Type 60, the values are as follows: 0.65 ± 0.04, 0.63 ± 0.04, 0.61 ± 0.06 and 0.61 ± 0.08. It is noted that the ratio slightly decreases with increasing holding time, and the standard deviation doubles. This result hints to the possibility that longer holding times allow for both borderline elastic-plastic and more elastic-dominated events to occur. A detailed analysis of elastic and plastic energies is presented in **Fig. 3b**, where the average values and standard deviations of the calculated elastic and plastic energies for each type of load function are plotted against the holing time. Elastic energies are represented by diamond markers, while plastic energies are represented by circular markers. Elastic work exhibits consistent values across diverse load functions, while the average value and standard deviation of plastic work increase with longer holding times. These results can be interpreted in the context of standard models for deformation under an indenter. Whether an expanding cavity model [28,29] or results from finite element methods [30]are used, the volume beneath the indenter can be divided into a hemispherical-like plastic zone, extending to a length $r$ equal to the contact radius $a$, and a broader elastic volume for distances greater than the contact radius. In the case of the amorphous structure of MGs, the confined and limited volume of the plastic zone may include both stable and unstable regions. Previous molecular dynamics studies [31,32] have shown that these regions correspond to structures such as Frank-Kasper and icosahedral motifs, respectively, which atomic-scale arrangements have nanometric dimensions. In contrast, the theoretically infinite and larger volume involved in elastic deformation is still a highly heterogeneous structure with nanometric components but exhibiting a higher degree of homogeneity on a larger, eventually micrometric, scale. In this work, time-independent elastic energies are believed to arise from the broader possibilities for energy dissipation within the extended elastic volume. Meanwhile, the increasing standard deviations of plastic work are attributed to the stochastic nanometric-scale heterogeneities within the limited plastic zone. It is worth noting that these time-dependent characteristics of plastic work are more clearly observed due to the indentation tests being conducted at low loads, well below the millinewton ranges typically used in broader hardness and mechanical property assessments. As a result, the plastic zone is smaller in our tests, enabling plastic effects arising from stochastic structural heterogeneity to be more readily observed. The increasing average values of plastic work are attributed to local configurational energy changes induced by the application of a constant load for a certain period, as the occurrence of pre-yielding plastic activity has already been documented in uniaxial elasto-static compression testing [33,34].

### *3.3 Clustering and creep analysis*

To further investigate the location dependence of nanoindentation-induced deformation resulting from different holding times, inspired by the work done on non-amorphous metallic alloys to correlate mechanical properties to microstructure leveraging ML [12–14], data are analyzed using an unsupervised expectation-maximization clustering algorithm: the GMM with a full covariance matrix. In this context, the term *cluster* refers to statistically defined groupings of data points with similar features. It does not correspond to any physical morphology or topological domain within the sample. The model is implemented in Python Scikit-learn library [35] to cluster the data based on measured hardness and creep displacement. It is noted that the creep response can be influenced by events occurring during the preceding loading segment. However, the investigation of creep behavior requires a prior loading phase. The 300 µN peak load used in this study contributes to limit the impact of loading-induced effects, allowing meaningful analysis of the creep regime. Hence, the creep and hardness data are analyzed within this experimental setup. Unlike studies on non-amorphous metallic alloys, where the number of Gaussian components is typically determined by the number of phases or features of interest identified through imaging techniques such as scanning electron microscopy and electron backscatter diffraction and confirmed by numerical estimation using information criteria (e.g. BIC, AIC), this work employs the forementioned criteria to numerically determine the optimal number of Gaussian distributions. This approach is motivated by the distinct nature of BMGs compared to multiphase crystalline metallic alloys, and the uncertainty surrounding the 'dual phase' simplified model of strongly versus weakly bonded regions. It allows for the inclusion of intermediate regions that can be represented and discussed. The optimal number of components is identified as the minimum value for each criterion, which, as shown in **Fig. 4a**, is equal to 3. The GMM is then fitted to the data, and each data point is assigned to a specific cluster based on its probability distribution. The clustered data are shown in **Fig. 4b**, where each dot represents an individual test. Side plots display the probability distributions for each cluster, with the solid line representing the kernel density estimate, which reflects the expected shape of the data without assuming any specific distribution function or normality. A bar plot in **Fig. 4c** illustrates the percentage of tests within each cluster, categorized by load function. As the holding time increases, the number of tests assigned to clusters corresponding to larger creep displacements also rises. Detailed plots for Types 10, 30, and 60 are provided in **Fig. 4d, 4e, and 4f**, respectively, and further information on the average coordinates of each cluster for each Type is available in the supplementary information.

Due to the larger number of data points available in each cluster for the 60-second holding time, including *clusters 0 and 1,* that are less populated in the other load function types, a detailed investigation of creep behavior is conducted for the *Type 60* dataset. First, the spatial distribution of the cluster numbers is shown on the map in **Fig. 5a**. A representative curve for each cluster group is then generated by averaging all tests within the respective cluster. In Fig. 5b, the data for the holding segment are plotted as a function of relative depth and time, with the depth set to zero at the start of the holding segment, which also corresponds to time zero. Curves are fitted using the following equation

$$h(t) = h_0 + at + b(t - t_0)^c , \qquad (1$$

where $h_0$ and $t_0$ are equal to zero and $a$, $b$ and $c$ are fitting parameters. The resulting values for the fitting parameters are provided in the legend of **Fig. 5b**. It is observed that the linear term $a$ is present only in *cluster 0*, indicating that the time evolution in this cluster exhibits a degree of linear dependence. In contrast, parameter *b*, which governs the non-linear time dependency, remains approximately constant across all curves. On the other hand, the exponent *c*, which characterizes the rate of time-dependence, shows a distinct trend, increasing from the shallower depths of *cluster 2* to the deeper depths of *cluster 0*. This variation in the exponent suggests that the local structural configuration plays a significant role in determining the nature of time-dependent energy dissipation, with regions of deeper creep displacement exhibiting a more pronounced time dependency. This behavior may be attributed to factors such as increased material heterogeneity with varying local atomic bonding state before and during the nanoindentation test, which could facilitate the activation of multiple percolation paths for energy dissipation. The different extent of the induced plastic zones for each cluster can be visualized by calculating the pressure distribution, as the pressure applied by a conical indenter over a contact radius *a* on the surface of a linearly elastic semi-infinite half-space [36]. The Berkovich indenter can in fact be approximated to a conical indenter with a 70.3-degree half-angle. The normalized contact pressure distribution $\frac{\sigma_z}{p_m}$ is expressed as

$$\frac{\sigma_z}{p_m} = -cosh^{-1}\frac{a}{r},  \quad (2$$

where *r* is the distance from the indenter-sample contact point. The contact radius $a_c$ is calculated from the contact depth as

$$a_c = 2\sqrt{3}h_c \tan 65.27. \quad (3$$

Normalized contact pressure distribution as a function of distance *r* is visualized in **Fig. 6a**, each contour is representative of a cluster with the maximum and minimum contact radii *a* corresponding to *cluster 0* and *2*, respectively. It is noted that the range of the distance *r* is between 80 and 150 nm, which is consistent with previously reported elastic correlation lengths in BMG of the same composition [5,8]. This suggests that, in addition to the elastic response being influenced by the microstructure, local configurations also play a significant role in small-scale plasticity. Based on the results from molecular dynamics simulations [31,32], the measurements falling into *cluster 0* could be associated with a weaker, more deformation-prone microstructure, potentially rich in icosahedral motifs. In contrast, *cluster 2* would correspond to regions rich in more stable and deformation-resistant Frank-Kasper structures. *Cluster 1*, on the other hand, would represent intermediate configurations, where neither stiffer nor more easily deformable motifs dominate. Elastic and plastic energies for each cluster are shown in **Fig. 6b**, following the same color coding introduced in previous figures. It is observed that the elastic contribution remains nearly constant across all tests, while the plastic energy varies distinctly between clusters. *Cluster 0* shows the largest plastic energy contribution, which decreases in *clusters 1* and *2.*

### 3.3.1. CCDF fitting and potential deformation mechanisms

The complementary cumulative distribution functions (CCDF) of plastic energy contributions is analyzed, as shown in **Fig. 7a and 7b**, to further investigate whether the different clusters correspond to distinct deformation mechanisms. In BMGs, plasticity is governed by shear transformation zones (STZs), which are associated with serrated flow. Both Gaussian and Weibull distribution models have been used to characterize the onset of plastic behavior [7,8]. Fitting equations and parameters are shown in the legend of each plot.

Figures 7a and 7b both present the CCDF of plastic energy contributions using the same dataset. Each figure illustrates the fitting of the data to a different model: a Gaussian distribution in Figure 7a and a Weibull distribution in Figure 7b. Figure 7a shows the fitting of all cluster data to the Gaussian model. The Gaussian distribution fits the events in *cluster 2* with an $R^2$ value of 0.9905. This fit is characteristic of events that exhibit small, random fluctuations around a mean value, with a relatively constant probability of occurrence over time. In the context of BMGs, this behavior can be associated with the activation of individual STZs, whose frequency of occurrence is primarily governed by the enthalpy of formation between the constituent elements of the alloy. Figure 7b shows the fitting of all data to the Weibull model. In contrast to the Gaussian distribution, the Weibull distribution provides the best fit for the events in *cluster 1*, with an $R^2$ value of 0.9744. *Cluster 0* exhibits good fitting to both Gaussian and Weibull distributions, with $R^2$ equal to 0.9836 and 0.9887, respectively. Given the similarity of the $R^2$ values, this metric alone is insufficient for distinguishing between the two models. Hence, additional evaluation criteria were used: the log-likelihood, AIC and BIC consistently indicated that the Weibull model provided a better fit to the data. Specifically, the Weibull model had a log-likelihood of -46.44, an AIC of 96.88, and a BIC of 102.39, whereas the Gaussian model yielded a log-likelihood of -67.00, an AIC of 138.00, and a BIC of 143.51. In the Weibull model, the probability of event occurrence evolves over time, capturing time-dependent or threshold-driven behavior. The model's two parameters, *k* and *λ*, offer insights into the underlying physical dynamics: a higher *k* value indicates that events are narrowly distributed around a specific condition, while a higher *λ* value corresponds to a greater energy threshold required for the initiation of an event. In the context of BMGs, this behavior might be indicative of the cooperative activation of STZs, which occurs once a well-defined energy barrier is overcome. **Figures 7c and 7d** represent a schematic of the proposed deformation mechanisms for individual and cooperative STZs activation, respectively.

4. *Conclusions*

To conclude, an annealed sample of $Zr_{50}Cu_{40}Al_{10}$ bulk metallic glass was investigated through nanoindentation testing. Four load functions were used, all with a peak load equal to 300 µN and same symmetrical loading rates but differing for the holding time, equal to 0, 10, 30 and 60 seconds. Indentation elastic and plastic energies and creep are investigated, leveraging machine learning clustering algorithms. It was found that the investigated BMG exhibits spatially heterogeneous and time-sensitive plastic behavior:

- Elastic energy contribution is consistent irrespective of holding time, while plastic energy average value and standard deviation increase with increasing holding time.
- Machine learning clustering with respect to hardness and creep displacement identifies three distinct groups. Each cluster shows similar elastic contributions, but distinct plastic energies, with the largest plastic energy corresponding to the cluster exhibiting the greatest creep displacements.
- Analysis of normalized contact pressure highlights the significant role of local configurations in small-scale plasticity within the nanometric range.
- Potential deformation mechanisms as individual and cooperative STZs activation are proposed based on the analysis of complementary cumulative distributions of the plastic energy contributions.

This work lays the foundation for further exploration of the small-scale behavior of BMG. Specifically, examining the effects of temperature on the observed deformation behavior will offer valuable insights into the role of viscous effects.

*Funding* T.O. and M.W. were supported by the Research Initiative of Structural Materials for Extreme Environment (RISME, No. JPMXP 1122684766) through the Ministry of Education, Culture, Sports, Science and Technology (MEXT), Japan.

*CRediT authorship contribution statement*

**Silvia Pomes**: Conceptualization, Data curation, Formal analysis, Investigation, Methodology, Writing – original draft, Writing – review & editing. **Takayuki Suzuki**: Resources. **Tomoya Enokizono:** Resources. **Nozomu Adachi**: Resources, Writing – review & editing. **Masato Wakeda**: Conceptualization, Funding acquisition, Writing – review & editing. **Takahito Ohmura**: Funding acquisition, Methodology, Resources, Supervision, Writing – review & editing.

*Declaration of competing interest*

The authors declare that they have no known competing financial interests or personal relationships that could have appeared to influence the work reported in this paper.

*Data availability*

The authors do not have permission to share data.

*Acknowledgement*

S.P. is grateful to Yamada Rui at Tohoku University for support with sample manufacturing; Hiroto Takanobu and Iwasaki Yutaka at NIMS for support with XRD and density measurements.

**Figures**

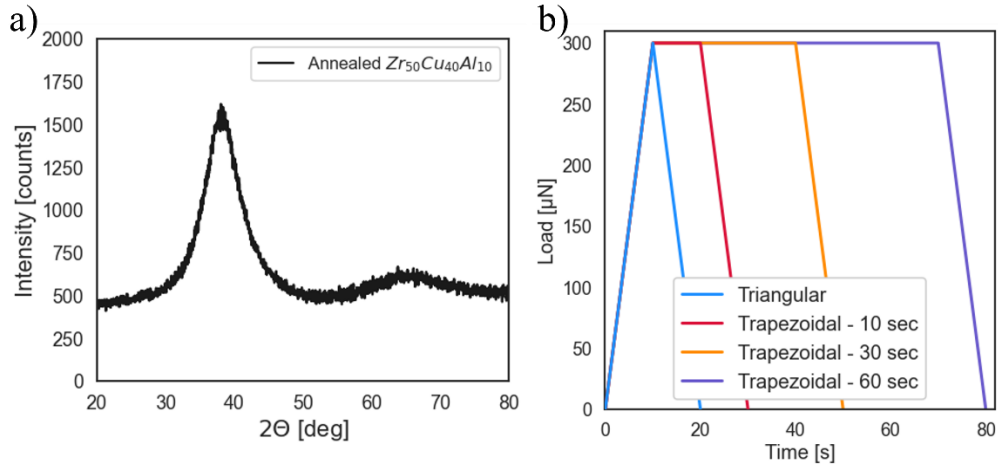

***Figure 1 a)*** *The amorphous structure of the annealed $Zr_{50}Cu_{40}Al_{10}$ bulk metallic glass sample is confirmed by XRD.* ***b)*** *The load functions used in quasi-static nanoindentation testing have the same peak load and symmetric loading rate but differ in their holding times, ranging from 0 seconds for the triangular load function to 10, 30, and 60 seconds for the trapezoidal load functions.*

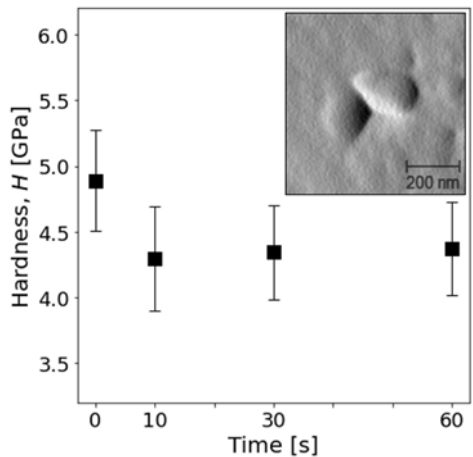

***Figure 2*** *Hardness values plotted against holding time show a higher value when the holding time is absent, while the hardness of the trapezoidal load functions remains comparable. Error bars represent the standard deviation and are consistent across all load functions. The inset shows the surface after the indentation test.*

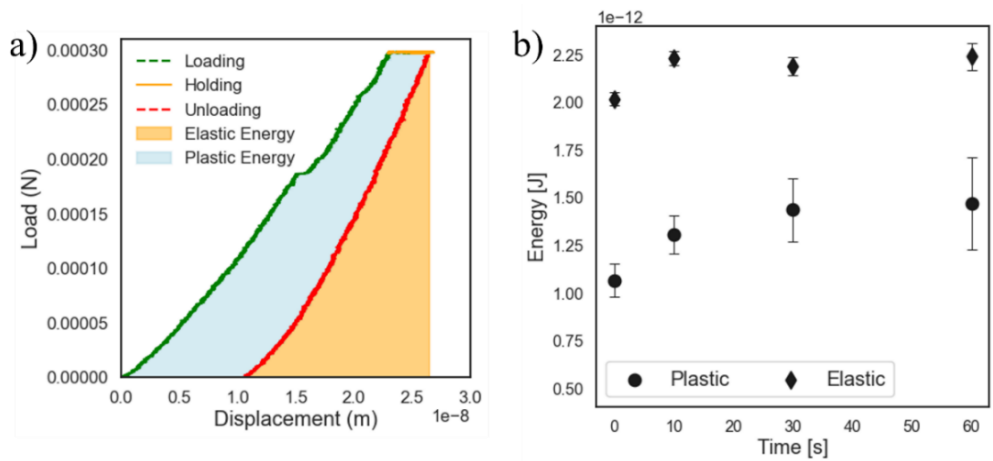

***Figure 3 a)*** *The elastic work is calculated as the area of yellow-hatched region underneath the unloading curve, while the plastic work is calculated as the area of blue-hatched region enclosed between the loading-holding-unloading segments.* ***b)***

*Elastic (diamond marker) and plastic (circular marker) energies plotted with respect to duration of holding segment. The elastic contribution is unaffected by the load function type, whereas the plastic contribution shows an increase in both the average value and standard deviation. It suggests that long-term deformation under a constant applied load is influenced by local microstructural features, in terms of energy dissipation.*

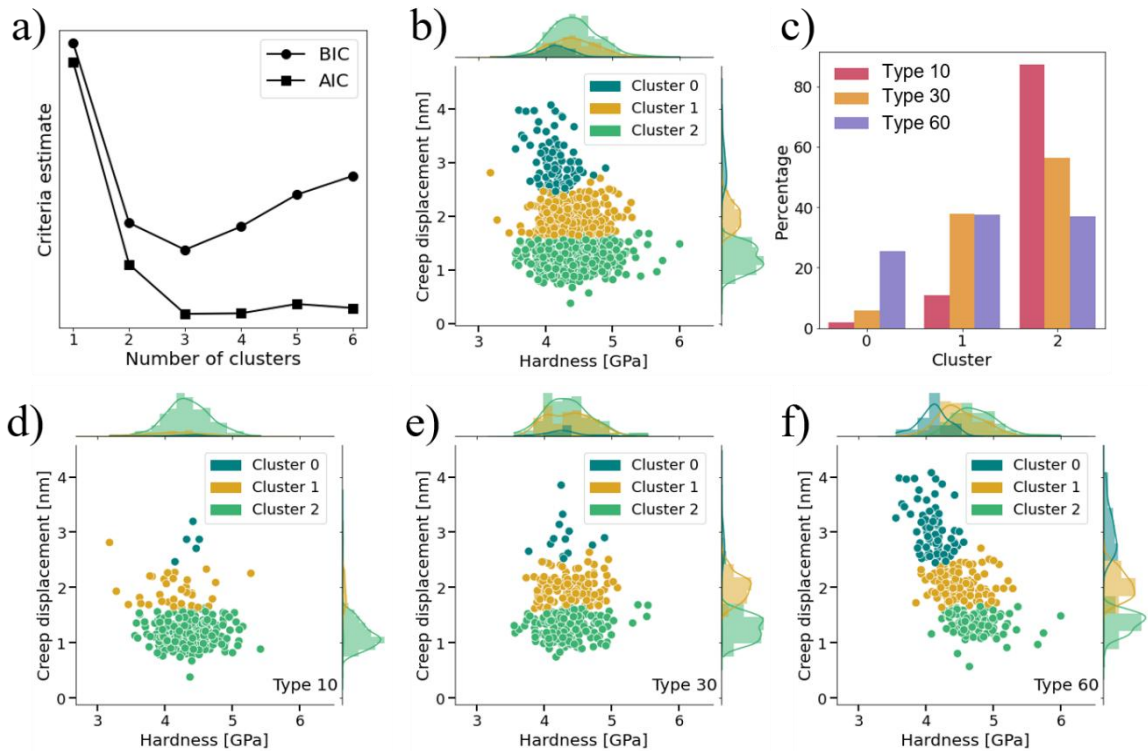

*Figure 4 Data are evaluated based on measured hardness and creep displacement, using a Gaussian Mixture Model clustering algorithm in Python: **a)** BIC and AIC estimation criteria suggest three as minimum ideal number of clusters. **b)** Clustered data from all tests conducted with a trapezoidal load function. Each dot on the main plot represents an individual test, while the distributions are shown in the side plots. **c)** A bar plot illustrating the percentage of tests in each cluster by load function. As the wait time increases, the number of tests in clusters corresponding to larger creep displacements also increases. Detailed plots for each holding time, namely 10, 30, and 60 seconds, are shown in panels **(d)**, **(e)**, and **(f)**, respectively.*

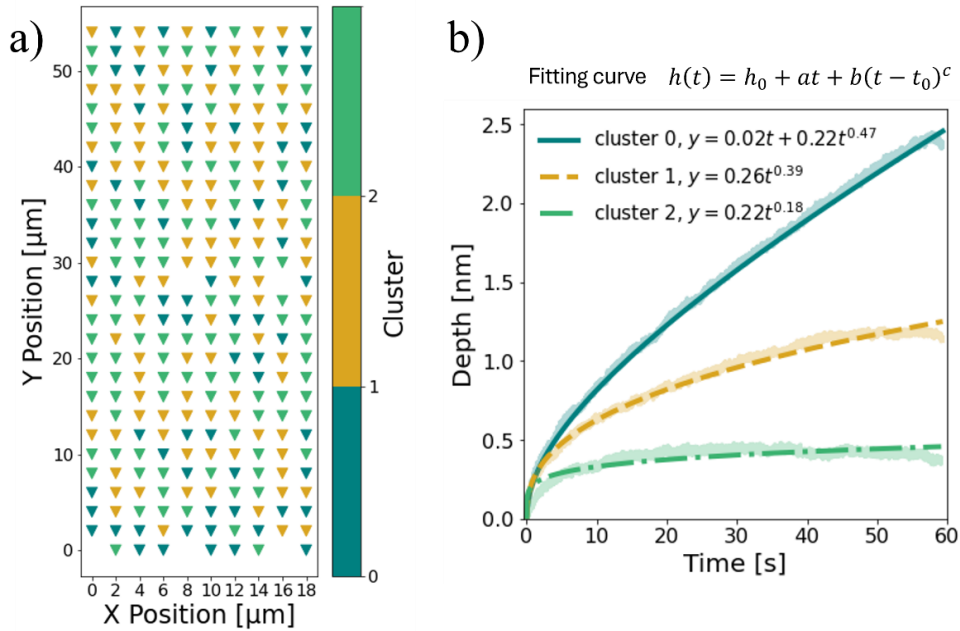

*Figure 5* Measurements collected using a load function with a 60-second holding time are analyzed. **a)** Spatial representation of cluster distribution. **b)** Creep fitting of the average representative curve for each cluster. The linear term $a$ is present only in cluster 0, while the non-linear term b is consistent across all clusters. The exponent of time-dependency c increases from the shallow depths in cluster 2 to the deeper depths in cluster 0, indicating a stronger time-dependency of energy dissipation mechanisms in cluster 0 configurations.

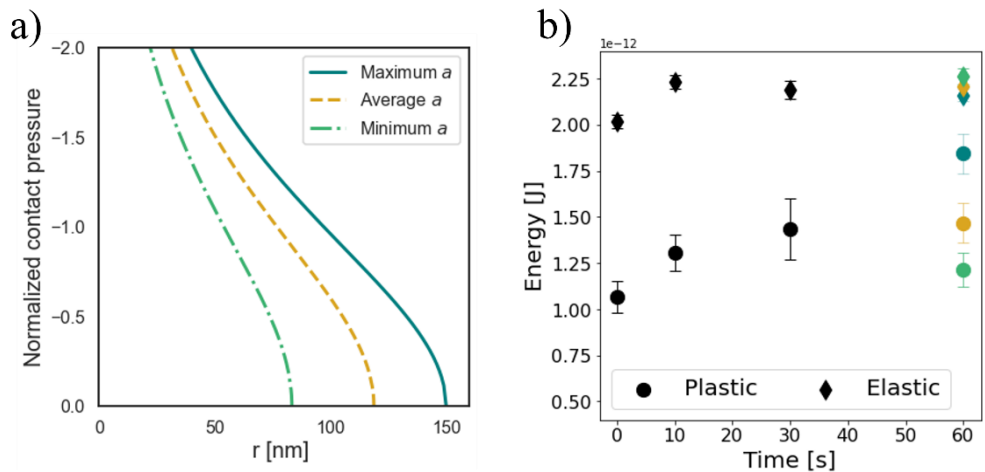

*Figure 6* **a)** Normalized contact pressure distribution as function of r, the distance from the indenter contact point. The pressure is zero when r equals a. **b)** Elastic and plastic energies contributions with respect to time, data for the 60-seconds holding load function are presented according to the clustering results.

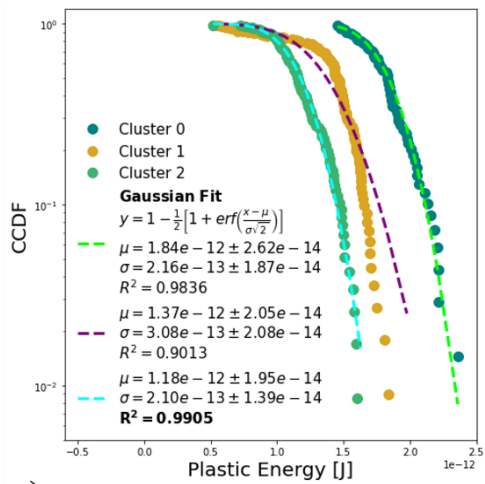 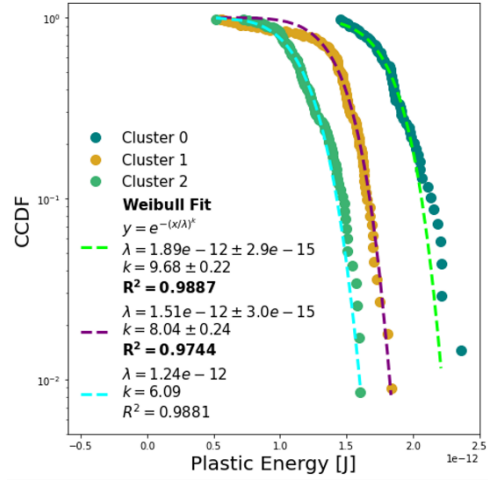

c) d)

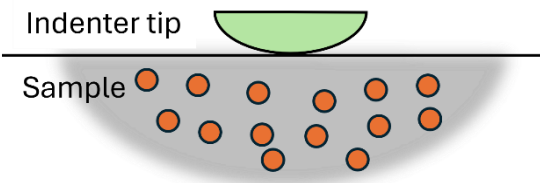 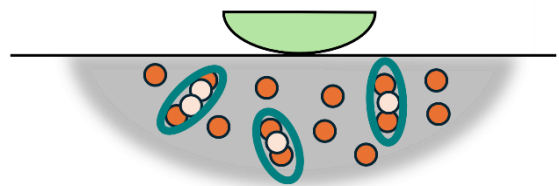

**Individual STZs activation**
Follows Gaussian distribution
Predominant in *cluster 2*

**Cooperative STZs re-arrangement**
Follows Weibull distribution
Predominant in *cluster 0*

*Figure 7* Complementary cumulative distribution function of plastic energy contributions of data collected with 60-second holding load function, according to clustering results. **a)** Gaussian distribution shows a good fit ($R^2$=0.9905) to cluster 2 data, while **b)** Weibull distribution best fits clusters 0 and 1 ($R^2$ as 0.9887 and 0.9744, respectively), highlighting the exponential or avalanche-like deformation dynamics. It is noted that cluster 0 exhibits satisfactory $R^2$-values for both fitting model: the goodness of fitting to the Weibull distribution was further confirmed by log-likelihood, AIC, and BIC as detailed in section 3.3.1 of the main text. Potential deformation mechanisms are proposed as **c)** individual STZs activation, predominant in cluster 2 and following a Gaussian distribution, and **d)** cooperative STZs re-arrangement, predominant in cluster 0 and following a Weibull distribution.

**Supplementary material**

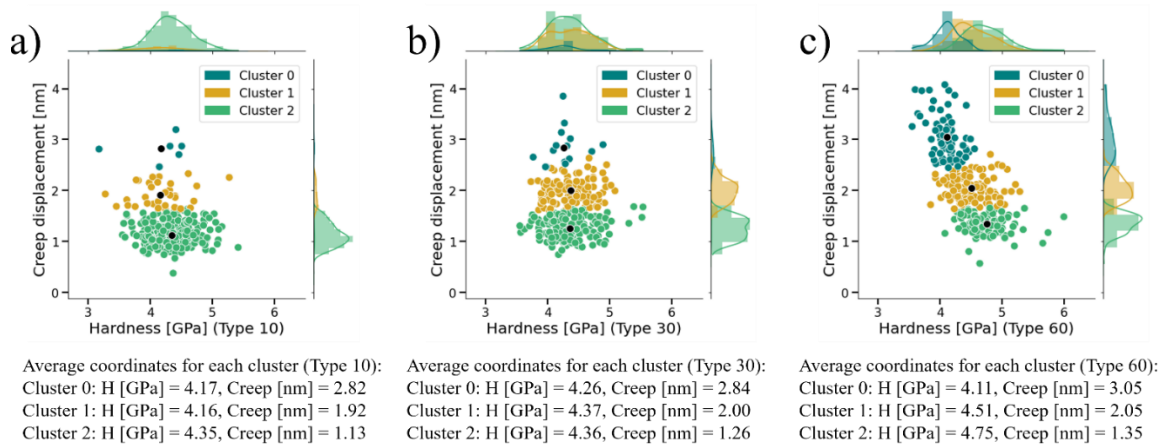

Figure S 1 Average coordinates for each cluster and a) Type 10, b) Type 30 and c) Type 60

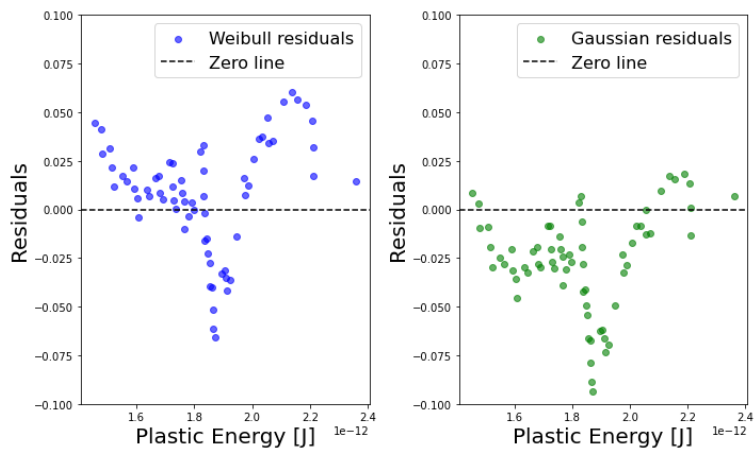

*Figure S 2 Residuals analysis to evaluate cluster 0 best fit revealed that the Weibull model produced symmetric residuals centered around zero, suggesting that it captured the underlying distribution of the data more accurately. In contrast, the residuals for the Gaussian model displayed asymmetry, further supporting the superiority of the Weibull model in representing the data.*